\documentclass[10pt, conference]{IEEEtran}
\IEEEoverridecommandlockouts
\usepackage{cite}
\usepackage{amsmath,amssymb,amsfonts}
\usepackage{algorithmic}
\usepackage{graphicx}
\usepackage{textcomp}
\usepackage{xcolor}
\usepackage{array}
\usepackage{url}

\usepackage{geometry}
\def\BibTeX{{\rm B\kern-.05em{\sc i\kern-.025em b}\kern-.08em
    T\kern-.1667em\lower.7ex\hbox{E}\kern-.125emX}}

\newcommand{\gametitle}{\textit{Star Collection}}
\newcommand{\toolname}{\textit{BiFuzz}}
\newcommand{\figref}[1]{Fig. \ref{#1}}

\begin{document}

\title{BiFuzz: A Two-Stage Fuzzing Tool\\for Open-World Video Games}

\author{\IEEEauthorblockN{Yusaku Kato}
\IEEEauthorblockA{\textit{Ritsumeikan University}\\
Ibaraki, Japan \\
is0610ks@ed.ritsumei.ac.jp}
\and
\IEEEauthorblockN{Norihiro Yoshida}
\IEEEauthorblockA{\textit{Ritsumeikan University}\\
Ibaraki, Japan \\
norihiro@fc.ritsumei.ac.jp}
\and
\IEEEauthorblockN{Erina Makihara}
\IEEEauthorblockA{\textit{Ritsumeikan University}\\
Ibaraki, Japan \\
makihara@fc.ritsumei.ac.jp}
\and
\IEEEauthorblockN{Katsuro Inoue}
\IEEEauthorblockA{\textit{Ritsumeikan University}\\
Ibaraki, Japan \\
inoue-k@fc.ritsumei.ac.jp}
}

\maketitle

\begin{abstract}
Open-world video games present a broader search space than other video games, posing challenges for test automation. Fuzzing, which generates new inputs by mutating an initial input, is commonly used to uncover issues. In this study, we proposed BiFuzz, a two-stage fuzzer designed for automated testing of open-world video games, and investigated its effectiveness. The results revealed that BiFuzz mutated the overall strategy of gameplay and test cases, including actual movement paths, step by step. Consequently, BiFuzz can detect character stuck issues. The tool and its video are at \url{https://github.com/Yusaku-Kato/BiFuzz} and  \url{ https://www.youtube.com/watch?v=VOrHfnLJSbk}.
\end{abstract}


\section{Introduction}
Increasing costs and prolonged development periods are significant problems in video game development\cite{Untitled27:online}\cite{2023shit96:online}. Automation is a way to address prolonged development periods\cite{5387726}. In video game development, several studies have been performed on automated testing\cite{prasetya2020navigation}\cite{10.1145/3550355.3552440}. Several video game companies employ scripts to simulate player actions to test automation\cite{murphy2014cowboys}. However, many developers remain reluctant to adopt test automation, as scripted tests are vulnerable to frequent software changes\cite{murphy2014cowboys}. Video games undergo frequent modifications throughout development, often driven by the game designers’ evolving subjective and artistic requirements. These changes aim to ensure the game remains enjoyable\cite{murphy2014cowboys}. Design-driven modifications can render automated tests obsolete, requiring continual updates to test content. Additionally, the cost of employing human testers for playtesting is currently lower than that of implementing test automation\cite{murphy2014cowboys}. For video game test automation, the above problems need to be solved.

It is challenging to automate testing, especially for open-world video games, which allow players to explore vast environments freely. In these games, players enjoy considerable autonomy in selecting from a complex system of mechanics, explorable spaces, and goals. The games feature a diverse array of objects (e.g., buildings and creatures) within a single scene. Furthermore, flexible movement mechanics and diverse object interactions enhance player freedom, contributing to an enormous state space that is challenging to explore\cite{10.1145/3102071.3102098}. 
The existing studies on automated playing in open-world video games have proposed LLM-based approaches\cite{wang2023voyager}. However, these approaches require large volumes of training data. As video games undergo frequent modifications throughout development, training data must be collected after each change. Thus, employing LLMs in video game development is cost-prohibitive and impractical.

To detect potential issues, the fuzzing testing approach repeatedly executes a program with automatically generated inputs that are often syntactically or semantically unexpected \cite{8863940}. Fuzzing requires no training data and adapts easily to frequent software changes. Furthermore, fuzzing can reduce development effort by eliminating the processes in which testers write scripts for each test case or operate the game. Accordingly, this study focuses on fuzzing for automated testing.

This paper presents \toolname, a two-stage fuzzing tool designed specifically to detect character stuck issues (CSI), which are common in open-world video games.
Our contributions are that:
\begin{itemize}
\item \toolname{} realizes a novel two-stage fuzzing approach that considers state transitions in open-world video games.
\item We demonstrate that \toolname{} can successfully detect pre-inserted CSIs when applied to an open-world video game developed using \textit{Unity}.
\end{itemize}



\begin{figure*}[ht] 
\begin{center} \includegraphics[width=0.7\textwidth]{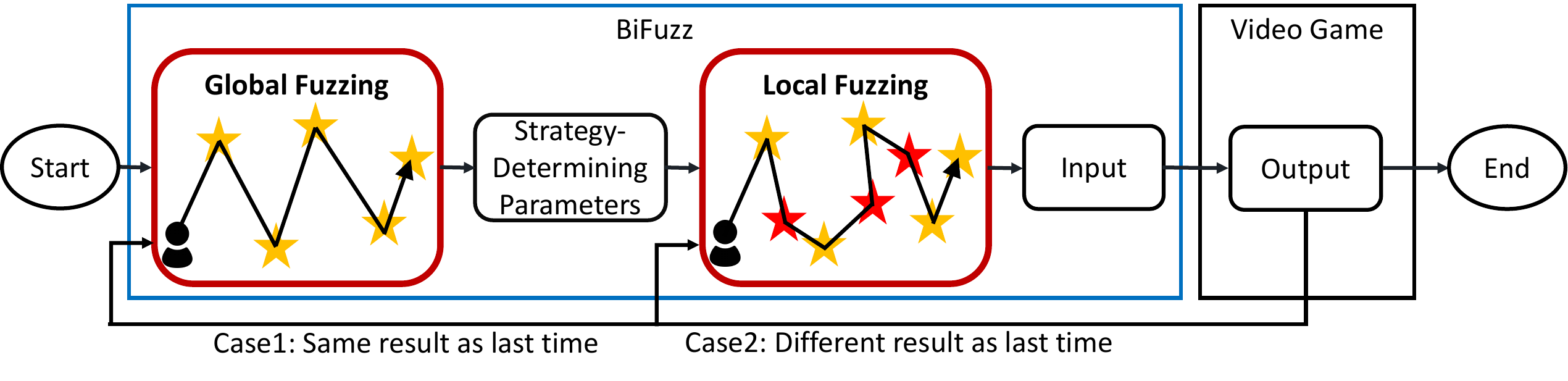} \caption{An overview of the fuzzing process in BiFuzz} \label{fig:method2} 
\end{center} 
\vspace{-8mm} 
\end{figure*}

\section{Related Work} \label{sec:related}
Fuzzing is an automated testing approach in which an initial input is repeatedly mutated to generate new inputs, termed fuzz, that are likely to cause issues. In this study, a `character' refers to an object controlled by the player in video games. ‘Fuzz’ denotes an input sequence used to move a character that is likely to cause issues. 
\textit{AFL}\footnote{\url{https://lcamtuf.coredump.cx/afl/}} is a representative coverage-guided grey-box fuzzer with a proven track record of detecting issues in various software. It measures code coverage for each input and uses this information to guide the generation of subsequent fuzz. Aschermann et al.\cite{aschermann2020IJON} proposed \textit{IJON} to enhance \textit{AFL} and applied it to Super Mario Bros. They demonstrated that conventional fuzzing based on code coverage performs inadequately for software with complex state transitions, such as video games. Furthermore, they proposed an extended tool named \textit{IJON} to improve \textit{AFL} based on annotations. Consequently, they demonstrated the alleviation of state explosion.

Modern video games released are typically developed using game engines \cite{SteamDB}. Game engines such as \textit{Unity} provide functions, e.g., object placement, physics calculations, and graphics drawing. Open-world video games are predominantly developed using such engines. Classical video games (e.g., Super Mario Bros.) have a small number of state variables, making them easy to identify and annotate for \textit{IJON} to work. Conversely, open-world video games developed using game engines have a larger number of state variables than those in classical video games, and these variables are difficult for testers to identify. This problem is common across game engines. Therefore, it is challenging to apply \textit{IJON} to the open-world video games.



\section{Two-Stage Fuzzing Tool: BiFuzz} 
We developed \toolname, which uses keypoints on the map to narrow the search space without annotations. In open-world video games, a player has various in-game destinations, or keypoints, and can freely change the routes for the keypoints. This flexibility contributes to a state explosion. To explore this state space efficiently, \toolname{} implements a two-stage fuzzing process. According to SteamDB\cite{SteamDB}, \textit{Unity} is the most widely used game engine. Therefore, we developed \toolname{} to operate on the \textit{Unity} platform. We developed it as a component within \textit{Unity}, allowing its scripts to be attached to a character object via drag-and-drop. This approach enables fuzzing for open-world video games developed using game engines incompatible with \textit{IJON}\cite{aschermann2020IJON}.

\figref{fig:method2} provides an overview of the \toolname{} fuzzing process. The tool operates within the IDE (i.e., \textit{Unity Editor}) and iteratively generates and mutates an initial input. \toolname{} performs fuzzing in two stages: \textit{Global Fuzzing (GF)} and \textit{Local Fuzzing (LF)}. \figref{fig:method2} includes both stages of fuzzing. In the \textit{GF} process, \toolname{} determines the order in which players pass through keypoints, which are represented as yellow stars. During \textit{LF}, waypoints, represented as red stars, are added between each pair of yellow stars determined in \textit{GF}. \toolname{} simulates player character movements by navigating through these waypoints, enabling detailed map exploration. In addition, execution traces and outcomes are recorded so that the generated inputs can be reproduced. \toolname{} repeats these two fuzzing stages until it processes the number of frames defined by a tester for the target video game. Let $m$ denote the number of iterations during the tester-defined time, and let $t$ be a natural number such that $t \leq m$. Let $R_1, R_2, ..., R_m$ be the iteration results obtained from the $1$st to the $m$-th iterations. Subsequently, each result $R_t$ and the corresponding detected issues $F_t$ are represented as $F_t = (R_t, C_t)$, where $R_t = Pass / Fail / Timeout$.
$C_t$ is defined only when $R_t$ is $Fail$, indicating the coordinate of an occurred issue. $R_t$ is $Fail$ if \toolname{} detects an issue, and $Pass$ if the entire input sequence is executed without detecting any issues. Furthermore, to prevent excessive delays caused by a large input sequence, the value of $R_t$ is set to $Timeout$ if the elapsed time of an iteration exceeds a tester-defined threshold.
The termination condition for \toolname{} is when a target video game reaches the number of frames specified by a tester. Furthermore, each iteration considers two cases. \toolname{} selects the appropriate case based on observing the $F_t$. The following outlines the two cases. If the result matches that of the previous iteration, \toolname{} proceeds to Case 1, executing both \textit{GF and LF}. Otherwise, it proceeds to Case 2, executing only \textit{LF}. The details of \textit{GF} and \textit{LF} are described below.

\subsection{Global Fuzzing} 
A character is an object controlled by the player. \toolname{} imitates typical player operations by determining the order in which the character passes through keypoints. This order is guided by a parameter that records each keypoint's pass/not pass status and priority. This paper defines a play style by the player operation characteristics expected by testers, and is configured prior to fuzzing. \toolname{} assigns weighted probabilities to candidate parameter values based on the specified play style. For example, in a play style where the character is expected to pass through all keypoints, all boolean parameters indicating keypoint passage are set to $true$. As illustrated, the parameter values are selected based on a weighted probability distribution.


\subsubsection*{\textbf{Strategy determination using parameters}}
\begin{figure}[t]
    \begin{center}
        \includegraphics[width=0.35\textwidth]{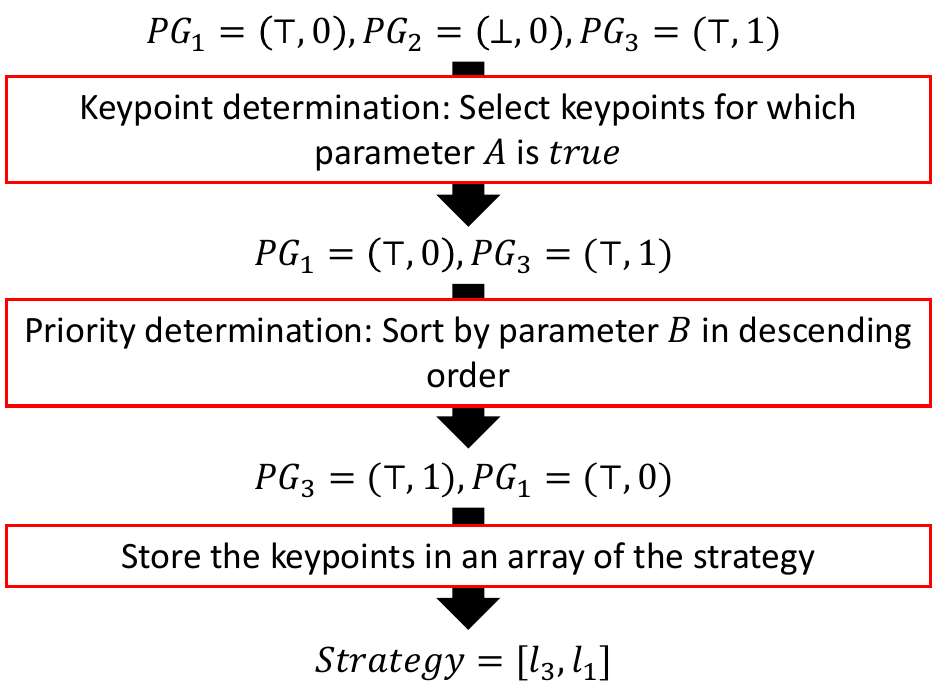}
        \caption{Strategy determination using parameters}
        \label{fig:strategy}
    \end{center}
    \vspace{-8mm} 
\end{figure}

In this study, a strategy refers to the keypoints a character passes through and the order in which they are traversed. This paragraph describes the process of determining the strategy. Let keypoints be stored in an array $k = [1, 2, ..., n]$, where $n$ is the number of keypoints. Let $x$ be a natural number such that $x \leq n$, Keypoint $x$ exists at coordinate $l_x$. \toolname{} employs strategy-determining parameters in this process. Let the parameters for each keypoint be denoted by $PG_x = (A_x, B_x)$. Here, $A_x$ is a boolean value that records the pass/not-pass status of Keypoint $x$, and $B_x$ is an integer between $0$ and $n$ that represents its priority. As an example, consider the case with Keypoints $1$, $2$, and $3$ in a game, and the parameters for each keypoint are $PG_1 = (\top, 0), PG_2 = (\bot, 0)$, and $PG_3 = (\top, 1)$. \figref{fig:strategy} shows the process of strategy determination under the above parameter settings. In this figure, as the parameter $A$ is $true$ for both $PG_1$ and $PG_3$ and $false$ for $PG_2$, the character aims for Keypoints $1$ and $3$. Furthermore, the parameter $B$, which indicates the priority level of Keypoint $3$, is $1$. Therefore, it is higher than Keypoint $1$, which is $0$. Consequently, the character aims for Keypoint $3$ before Keypoint $1$. Thus, the strategy in this figure becomes $Strategy = [l_3, l_1]$.

\subsection{Local Fuzzing} 
To add waypoints into the order determined in \textit{GF}, \toolname{} employs parameters in a consistent manner. Each parameter in \textit{LF} includes the number and coordinates of waypoints, and it is selected based on the probabilities weighted by the specified play style.

\subsubsection*{\textbf{Degree of freedom determination using parameters}}
\begin{figure}[t]
    \begin{center}
        \includegraphics[width=0.8\columnwidth]{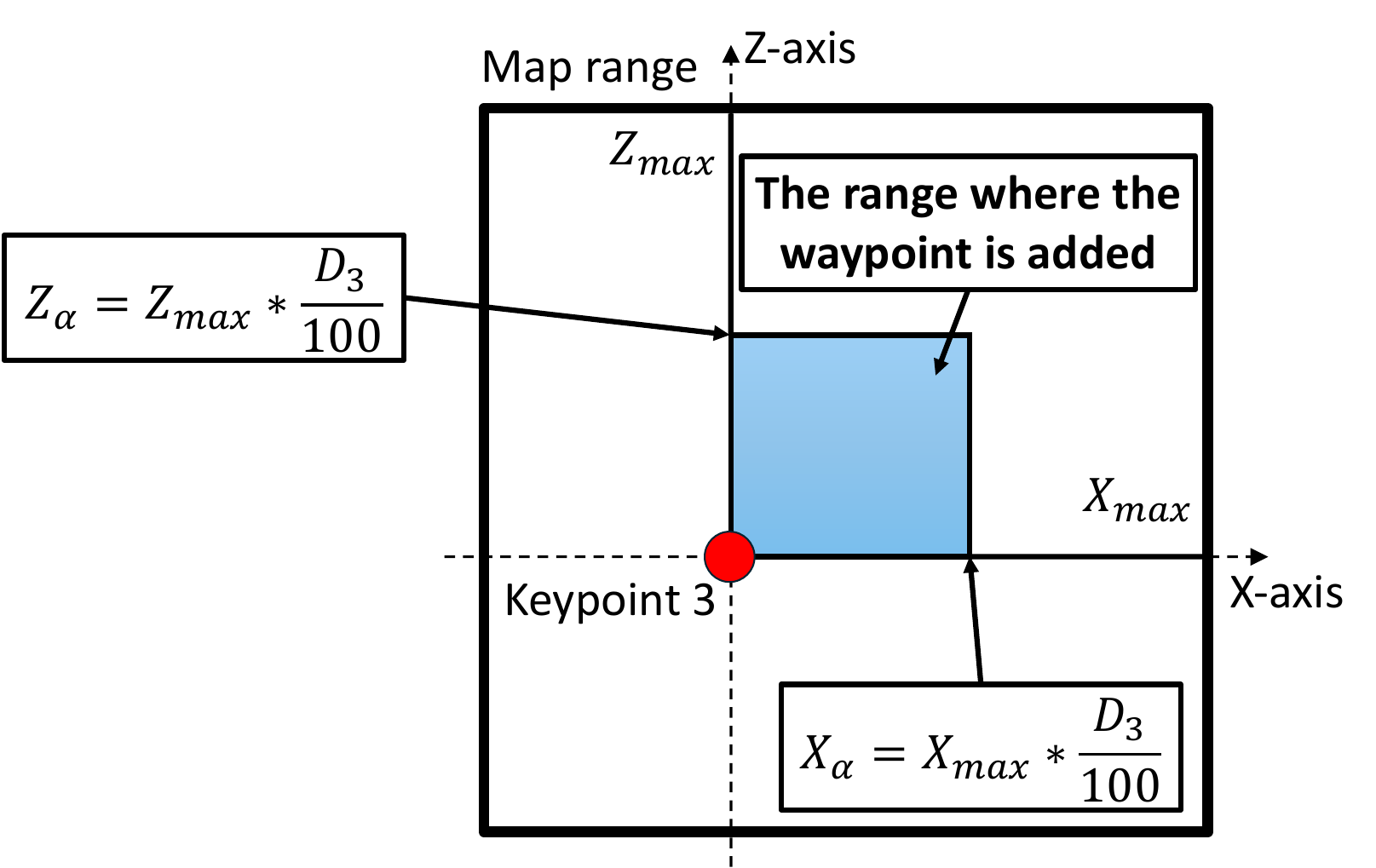}
        \caption{Degree of freedom determination using parameters}
        \label{fig:via}
    \end{center}
    \vspace{-8mm} 
\end{figure}

In this study, \toolname{} determines a degree of freedom by adding waypoints to a strategy. This paragraph explains the process of incorporating waypoints into the strategy determined in \textit{GF} using waypoint-adding parameters. Let parameters for each keypoint be denoted by $PL_x = (C_x, D_x, E_x)$. Here, $C_x$ is an integer between $0$ and $99$ that represents the number of waypoints to be added between Keypoint $x$ and the next Keypoint that a character passes through. $D_x$ is an integer between $0$ and $99$ that represents the maximum distance from Keypoint $x$ to a waypoint. Moreover, $E_x$ is an integer between $1$ and $4$ that determines the direction of the waypoint from Keypoint $x$. In the same \textit{GF} explanatory example, a new element is added to the array $Strategy = [l_3, l_1]$. As no keypoint exists after Keypoint $1$ in $Strategy$, only the parameters for $PL_3$ are used. Let $PL_3$ be defined as $PL_3 = (1, 50, 1)$. \figref{fig:via} shows a top-down view of the map in \gametitle{}, an open-world video game developed using \textit{Unity}. When $PL_3$ is set as above, an added waypoint is randomly selected from the light blue region shown in \figref{fig:via}.

Next, based on \figref{fig:via}, we explain how the degrees of freedom are determined using parameters. First, as the value of parameter $C_3$ is $1$, \toolname{} adds one waypoint. Moreover, from $E_3 = 1$, the added waypoint lies forward-right, assuming the positive Z-axis represents the front from the character's perspective. In addition, as $D_3 = 50$, the maximum distance from Keypoint $3$ to the waypoint is set to half ($50\%$) of the minimum distance from Keypoint $3$ to each side of the map. This approach enables \toolname{} to explore the state space more effectively.

\section{Evaluation}
Open-world video games generally feature distinct input types, such as simultaneous inputs and inputs of arbitrary length with arbitrary intervals. It is difficult to compare \textit{IJON}, the existing fuzzer for video games, with \toolname{} because it does not support these types of inputs. To this end, we evaluated the following experiments to confirm that \toolname{} performs effective testing. Specifically, we prepared two play styles for the fuzzing, conducted fuzzing with each, and confirmed how much the number of issues detected could be increased. 
\gametitle{} is an open-world video game developed by the first author using \textit{Unity}. We applied \toolname{} to one scenario in the game. The scenario was developed to confirm the effectiveness of \toolname{} for addressing the CSIs that commonly occur in open-world video games. The issue occurs when its movement is blocked due to colliding with the terrain or surrounding objects, hindering progress in the game.
In the game, a player operates a character in third-person view, with the objective of collecting items located at keypoints. Character movement is controlled using the W, A, S, and D keys, while the viewpoint is adjusted with the mouse.


We present the results of the test for the scenario that includes CSIs. First, we created a hundred objects for implementing the issues, and indexed them from 1 to 100. Then, we tested each of the Play styles A and B. In Play style A (PS-A), a character passes through fewer keypoints and waypoints than in Play style B (PS-B).
We applied \toolname{} to the game, and executed it on the IDE (i.e., \textit{Unity Editor}) for 2,592k frames (12 hrs.) with the target frame rate of 60 fps. We conducted this routine five times.



\begin{figure}[t]
    \begin{center}
        \includegraphics[width=0.45\textwidth]{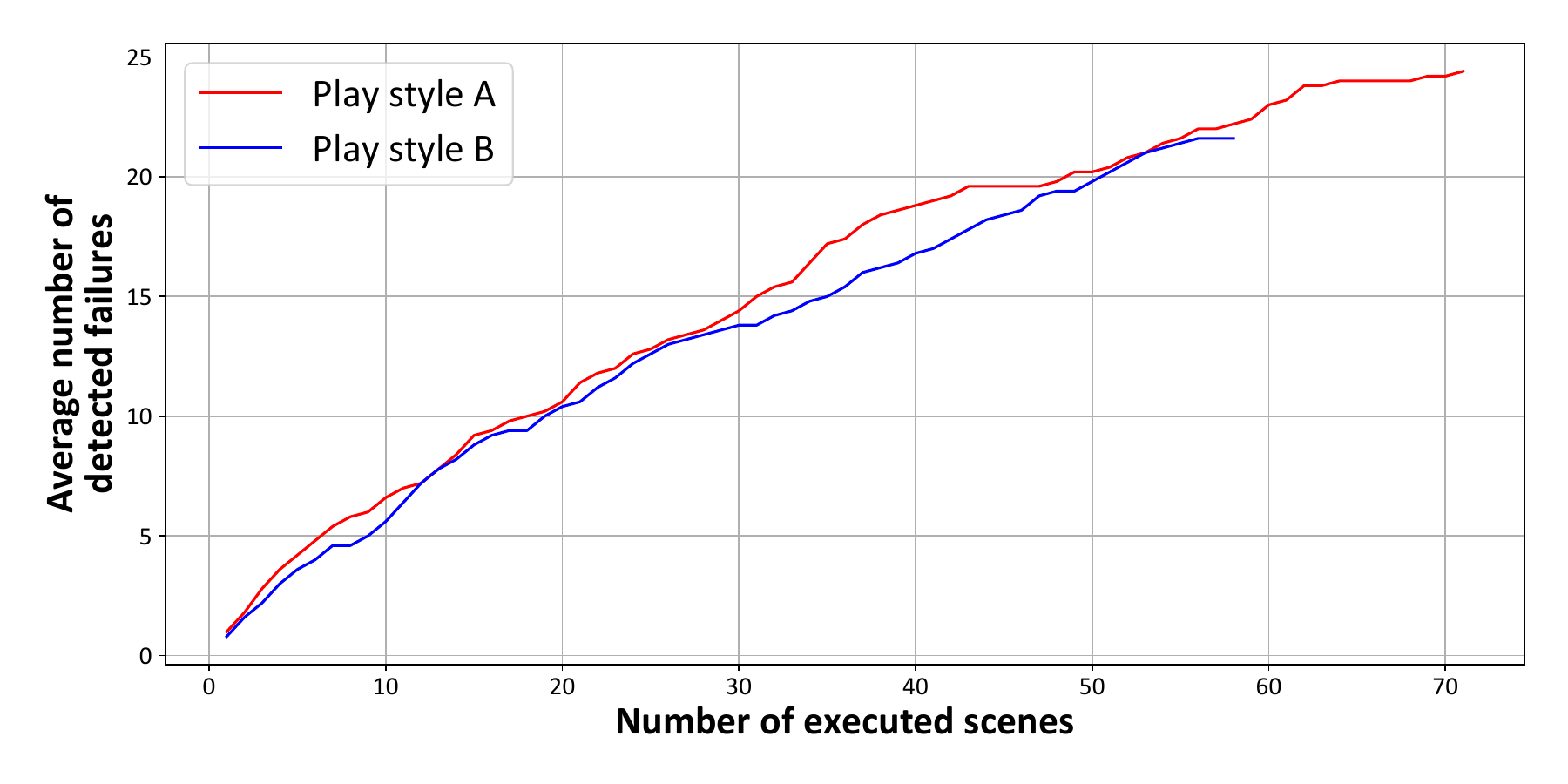}
        \caption{Number of detected issues}
        \label{fig:graph}
    \end{center}
    \vspace{-8mm} 
\end{figure}

On average, 8.4 more scenes were executed during testing with PS-A than with PS-B. This is because PS-A involves fewer keypoints and waypoints, resulting in a relatively smaller input in one scene. Additionally, the number of detected issues indicates how many of the 100 predefined issues in the scenario were identified. In PS-A, an average of 2.8 more issues were detected than in PS-B. \figref{fig:graph} presents a graph comparing the trends in the number of issues detected in PS-A and PS-B, with the average number detected on the vertical axis and the number of executed scenes on the horizontal axis. The graph shows that PS-A detected more than PS-B, primarily due to the higher number of scenes executed during testing. A total of 64 issues were detected using PS-A and 56 using PS-B. We analyzed the overlap and uniqueness of these issues across five runs. We found that, 43 issues were common to both styles, 21 were unique to PS-A, and 13 were unique to PS-B. 
Using different play styles enables the detection of distinct issues that may be missed by others.

\section{Conclusion}
We propose \toolname, a two-stage fuzzing tool designed specifically to detect CSIs.
The experimental results show that each of the two play styles detected unique issues. This indicates that \toolname{} can detect issues, and that performing fuzzing with two distinct play styles uncovers a greater number of CSIs.

However, the evaluation of \toolname{} was limited to a single custom game scenario, \gametitle{}, which makes the scale of the experiment insufficient. Therefore, in future work, we plan to establish baselines drawing on existing methods and conduct comparative evaluations with \toolname{}. Moreover, \toolname{} currently supports movement on the planar axes using the W, A, S, and D keys, and is unsupported for other input methods or three-dimensional movement. Nevertheless, it is technically feasible to extend the implementation to accommodate these in the future. In addition, since the experiments in this paper focused only on CSIs, future evaluations should also include other common video game issues, including graphics rendering errors and performance issues, to enhance the generality of the conclusions.

\section*{Acknowledgment}
This work was supported by JSPS KAKENHI Grant Number JP24K02923 and JST AIP Acceleration Research JPMJCR25U7, Japan.


\bibliographystyle{IEEEtran} 
\bibliography{reference} 

@misc{Untitled27:online,
  author = {{Competition \& Markets Authority}},
  title = {Anticipated acquisition by {Microsoft} of {Activision Blizzard, Inc.} Final report},
  url = "https://x.gd/sbSOK",
}

@misc{2023shit96:online,
  author = {Stephen Totilo},
  title = {Axios Gaming},
  url = {https://www.axios.com/newsletters/axios-gaming-57427f17-9caf-468f-bdca-d60a1ee2da09.html},
}

@inproceedings{murphy2014cowboys,
  author = {Murphy-Hill, E. and Zimmermann, T. and Nagappan, N.},
  title = {Cowboys, ankle sprains, and keepers of quality: how is video game development different from software development?},
  year = {2014},
  booktitle = {Proc. of ICSE},
  pages = {1–11}
}

@article{5387726,
  author = {Hoffnagle, G. F. and Beregi, W. E.},
  title = {Automating the software development process},
  year = {1985},
  volume = {24},
  number = {2},
  journal = {IBM Syst. J.},
  pages = {102–120}
}

@inproceedings{prasetya2020navigation,
  author = {Prasetya, I. S. W. B. and Voshol, M. and Tanis, T. and Smits, A. and Smit, B. and Mourik, J. v. and Klunder, M. and Hoogmoed, F. and Hinlopen, S. and Casteren, A. v. and Berg, J. v. d. and Prasetya, N. G. W. Y. and Shirzadehhajimahmood, S. and Ansari, S. G.},
  title = {Navigation and exploration in 3D-game automated play testing},
  year = {2020},
  booktitle = {Proc. of A-TEST},
  pages = {3–9}
}

@inproceedings{10.1145/3550355.3552440,
  author = {Casamayor, R. and Arcega, L. and P\'{e}rez, F. and Cetina, C.},
  title = {Bug localization in game software engineering: evolving simulations to locate bugs in software models of video games},
  year = {2022},
  booktitle = {in Proc. of MODELS},
  pages = {356–366}
}

@inproceedings{wang2023voyager,
  title={Voyager: An Open-Ended Embodied Agent with Large Language Models},
  author={Wang, G. and Xie, Y. and Jiang, Y. and Mandlekar, A. and Xiao, C. and Zhu, Y. and Fan, L. and Anandkumar, A.},
  booktitle={Proc. of NeurIPS},
  year={2023}
}

@misc{afl,
  author = {Zalewski, M.},
  title = {American Fuzzy Lop},
  howpublished = {\url{https://lcamtuf.coredump.cx/afl/}},
  month = {},
  year = {},
  note = "[Online; accessed 2025-01-05]"
}

@inproceedings{aschermann2020ijon,
  title={Ijon: Exploring deep state spaces via fuzzing},
  author={Aschermann, C. and Schumilo, S. and Abbasi, A. and Holz, T.},
  booktitle={Proc. of IEEE S\&P},
  pages={1597--1612},
  year={2020}
}

@misc{SteamDB,
  author = {SteamDB},
  title = {What are games built with and what technologies do they use?},
  howpublished = {\url{https://steamdb.info/tech/}},
  note = "[Online; accessed 2025-01-12]"
}

@ARTICLE{8863940,
  author={Manès, V. J. M. and Han, H. and Han, C. and Cha, S. K. and Egele, M. and Schwartz, E. J. and Woo, M.},
  journal={IEEE Trans. Softw. Eng.}, 
  title={The Art, Science, and Engineering of Fuzzing: A Survey}, 
  year={2021},
  volume={47},
  number={11},
  pages={2312-2331}
}

@inproceedings{10.1145/3102071.3102098,
author = {Alexander, Ryan and Martens, Chris},
title = {Deriving quests from open world mechanics},
year = {2017},
isbn = {9781450353199},
booktitle = {Proc. of FDG},
articleno = {12},
numpages = {7},
location = {Hyannis, Massachusetts},
}

\end{document}